\documentclass[prl,twocolumn,reprint,superscriptaddress,showpac]{revtex4-2}

\usepackage{feynmp,amsmath,amssymb,graphicx,subfigure,hyperref,bm,mathrsfs,slashed}
\usepackage{CJK}

\begin{document}

\title{Longitudinal optical conductivities of tilted Weyl fermion in arbitrary dimensionality}

\author{Jian-Tong Hou}
\affiliation{College of Physical Science and Technology and Center for Theoretical Physics, Sichuan University, Chengdu, Sichuan 610064, China}
\affiliation{Department of Physics, Institute of Solid State Physics and Center for Computational Sciences, Sichuan Normal University, Chengdu, Sichuan 610066, China}

\author{Peng Wang}
\affiliation{College of Physical Science and Technology and Center for Theoretical Physics, Sichuan University, Chengdu, Sichuan 610064, China}

\author{Hong Guo}
\thanks{guo@physics.mcgill.ca}
\affiliation{Department of Physics, McGill University, Montreal, Quebec H3A 2T8, Canada}
\affiliation{Department of Physics, Institute of Solid State Physics and Center for Computational Sciences,
Sichuan Normal University, Chengdu, Sichuan 610066, China}

\author{Hao-Ran Chang}
\thanks{hrchang@mail.ustc.edu.cn}
\affiliation{Department of Physics, Institute of Solid State Physics and Center for Computational Sciences, Sichuan Normal University, Chengdu, Sichuan 610066, China}
\affiliation{College of Physical Science and Technology and Center for Theoretical Physics, Sichuan University, Chengdu, Sichuan 610064, China}

\date{\today}
		
\begin{abstract}
The unified form of longitudinal optical conductivities (LOCs) in the tilted Weyl fermions for arbitrary spatial dimensionality are analytically calculated and expressed in terms of the joint density of state. The results are valid for both undoped and doped cases, both parallel and perpendicular components, and all the tilted phases. In addition, they reproduce analytical results of previous works for one-dimensional, two-dimensional, and three-dimensional tilted Weyl systems. The robust fixed point $\Gamma_{\chi}^{(\mathrm{IB})}(\omega=2\mu,d\ge 2;\mu,0<t\le 2)=1/2$ is universal for tilted Dirac bands in arbitrary spatial dimensionality. Our work provides not only a once-for-all method prior to the one-by-one calculation of the LOCs but also offer deep insights into the impacts of dimensionality in the tilted Weyl fermions.
\end{abstract}

\maketitle

The spatial dimensionality of systems and energy dispersion of carriers \cite{Vignale2005,Mahan2007} dominantly
determine most kinds of physical properties. A unified evaluation of physical properties in arbitrary spatial
dimensionality can not only provide a once-for-all method prior to the one-by-one calculation but also offer deep
insights into the impacts of dimensionality. For example, the Ruderman-Kittel-Kasuya-Yosida (RKKY) interaction \cite{RKKY3DEGPR1954,RKKY3DEGPRPK1956,RKKY3DEGPR1957} in arbitrary dimensionality \cite{Larsen1985,Aristov1997}
can automatically restore to it in one-dimensional (1D) \cite{RKKY12DEGLitvinov1998,RKKY1DVignale2005}, two-dimensional
(2D) \cite{RKKY2DEGKorenblit1975,RKKY2DEGFischer1975,RKKY2DEGBeal-Monod1987,RKKY12DEGLitvinov1998,RKKY1DVignale2005,
RKKY23DGiuliani2005}, and three-dimensional (3D) electron gases \cite{RKKY3DEGPR1954,RKKY3DEGPRPK1956,RKKY3DEGPR1957,
RKKY23DGiuliani2005}.

The low energy electronic structure of Weyl fermions in momentum space \cite{Science2004,RMP2009,RMP2018,SatoPRB2018,
Yangnjp2020,Zhunjp2020,WangACSNano2021,Zhangnjp2022,YueNanoLett2022} may be categorized into four distinct ``phases" via
a tilt parameter $t$, namely the untilted phase ($t=0$), type-I phase ($0<t<1$), type-II phase ($t>1$), and type-III phase ($t=1$) \cite{JPSJ2006,Zhou8Pmmn2014PRL,Science8Pmmn2015,Nature2015,Volovik2017,Volovik2018}. Earlier works on the longitudinal
optical conductivities (LOCs) focused mostly on the LOCs of untilted phase in 2D \cite{PRLCarbotte2006,PRBGusynin2007,
PRLMikhailov2007,PRLMarel2008,PRLMak2008,PRBStauber2008} and 3D \cite{PRBAshby2014}. Later, theoretical researches on the
LOCs have been extended to the tilted phases in 2D \cite{JPSJNishine2010,PRBVerma2017,PRBHerrera2019,PRBWild2022,PRBTan2022}
and 3D \cite{PRBCarbotte2016,PRBCarbotte2017}. Recently, our unified comprehensive theoretical analysis \cite{arXivHou2022} indicates that the spatial dimensionality and tilted phases play a significant role in determining the physical behaviors of
the LOCs. It is the purpose of this paper to offer the exact derivation of LOCs of tilted Weyl fermion in $(d+1)$-dimensional spacetime.

In $(d+1)$-dimensional spacetime, the Lagrangian density for tilted Weyl fermion $\psi$ interacting with electromagnetic field $A_{\nu}=(A_{0},A_{j})=(A^{0},-A^{j})=(\varphi,\boldsymbol{A})$ at finite chemical potential $\mu$ reads
\begin{widetext}
\begin{align}
\mathscr{L}=&\bar{\psi}\left[\left(\tau^{0}\otimes\gamma^{0}\right)~c~\mathrm{i}\hbar\partial_{0}
+\left(\tau^{0}\otimes\gamma^{j}\right)~v_{F}~\mathrm{i}\hbar\partial_{j}
+\left(\tau^{3}\otimes\gamma^{0}\right)~v_{t}~\mathrm{i}\hbar\partial_{1}
+\left(\tau^{0}\otimes\gamma^{0}\right)\mu\right] \psi
-\frac{1}{4}F_{\rho\sigma}F^{\rho\sigma}-\frac{1}{2}\left(\partial^{\nu}A_{\nu}\right)^2
\nonumber\\&
-\bar{\psi}\left[ e\left(\tau^{0}\otimes\gamma^{0}\right)A_{0}
+e\frac{v_{F}}{c}\left(\tau^{0}\otimes\gamma^{j}\right)A_{j}
+e\frac{v_{t}}{c}\left(\tau^{3}\otimes\gamma^{0}\right)A_{1}\right] \psi,
\end{align}
where the derivatives are defined as $\partial_{\nu}=\frac{\partial}{\partial x^{\nu}}$ and the Dirac matrices satisfy $\{\gamma^{a},\gamma^{b}\}=2g^{ab}$ with \ensuremath{g^{ab}=g_{ab}=\text{diag}(1,-1,\cdots,-1)}, \ensuremath{x_{\nu}=g_{\nu\rho}x^{\rho}} and $x^0=ct$, $\bar{\psi}=\psi^{\dagger}\left(\tau^{0}\otimes\gamma^{0}\right)$, $F_{\rho\sigma}=\partial_{\rho}A_{\sigma}-\partial_{\rho}A_{\nu}$ is the electromagnetic field tensor in $(d+1)$-dimensional Minkowski spacetime with $d$ the number of spatial dimensions.

The free Hamiltonian density at zero temperature and zero chemical potential takes
\begin{align}
\mathscr{H}_{0}&=\psi^{\dagger}\left[ v_{t}\left(\tau^{3}\otimes I\right)(-\mathrm{i}\hbar \partial_{1})
+v_{F}\left(\tau^{0}\otimes\gamma^{0}\boldsymbol{\gamma}\right)\cdot(-\mathrm{i}\hbar \boldsymbol{\nabla})\right] \psi
=\psi^{\dagger}\left[\left(\tau^{3}\otimes I\right) v_{t}~\hbar k^{1}
+v_{F}~\left(\tau^{0}\otimes \boldsymbol{\alpha}\right)\cdot(\hbar \boldsymbol{k})\right]\psi,
\end{align}
where Dirac matrices $\alpha^{j}=\gamma^{0}\gamma^{j}$, identical matrix $I=\gamma^{0}\gamma^{0}$, the wave vector $k=\left(k^{0},k^{j}\right)=\left(\frac{\Omega}{c},\boldsymbol{k}\right)$. This free Hamiltonian gives rise to a tilted energy eigenvalue along the first spatial component as
\begin{equation}\label{Eq.3}
\varepsilon_{\kappa}^{\lambda}(\boldsymbol{k})=\kappa v_{t}~\hbar k^{1}+\lambda v_{F}~\hbar|\boldsymbol{k}|
=v_{F}\left(\kappa t~\hbar k^{1}+\lambda ~\hbar|\boldsymbol{k}|\right),
\end{equation}
where $\kappa=\pm$ represents different valleys, $\lambda=\pm$ denotes different bands, $t=v_t/v_F$ characterizes the band \emph{tilting} along the first spatial direction, with $t=0$ for untilted phase,
$0<t<1$ for type-I phase, $t=1$ for type-III phase, and $t>1$ for type-II phase.
	
The photon self-energy in the $(d+1)$-dimensional Minkowski spacetime follows that
\begin{align}
\label{Eq.5}
\Pi^{\nu\rho}(\omega,\boldsymbol{q})=-\frac{\mathrm{i}}{\beta_{T}}\sum_{\Omega_{m}}
\int\frac{\text{d}^{d}\boldsymbol{k}}{(2\pi)^{d}}\mathrm{Tr}\left[\mathrm{i}V^{\nu}S_{F}(\mathrm{i}\Omega_{m};\boldsymbol{k})
\mathrm{i}V^{\rho}S_{F}(\mathrm{i}\Omega_{m}+\omega+\mathrm{i}\epsilon;\boldsymbol{k}+\boldsymbol{q})\right],
\end{align}
where $\epsilon$ is a positive infinitesimal, $\beta_{T}=1/(k_{B}T)$ with $k_{B}$ the Boltzmann constant, $V^{\rho}$ represents the interaction vertex
\begin{align}
V^{\rho}&=-e\left\{\left(\tau^{0}\otimes\gamma^{0}\right)~c~\delta_{0}^{\rho}
+v_{F}\left[\left(\tau^{3}\otimes\gamma^{0}\right)~t~\delta_{1}^{\rho}
+\left(\tau^{0}\otimes\gamma^{j}\right)\delta_{j}^{\rho}\right]\right\},
\end{align}
and the Green's function for tilted Weyl fermion with chemical potential $\mu$ is defined as
\begin{align}
S_{F}(k^{0},\boldsymbol{k})
&=\frac{\mathrm{i}}{\left(\tau^{0}\otimes\gamma^{0}\right)\left(\mathrm{i}\hbar\Omega_{m}-\mu\right)
-v_{F}~\left[\left(\tau^{3}\otimes\gamma^{0}\right)t~\hbar k^{1}
+\left(\tau^{0}\otimes\gamma^{j}\right)~\hbar k^{j}\right]}
=S_{F}(\mathrm{i}\Omega_{m},\boldsymbol{k}),
\end{align}
with the frequencies related to Matasubara sum $\Omega_{m}=(2m+1)\pi/\beta_{T}$ for $n=0,\pm1,\pm2,\cdots$. For convenience, we set $\hbar=c=k_B=1$.

The real part of longitudinal optical conductivities (LOCs) defined by
\begin{align}
\mathrm{Re}\sigma_{jj}(\omega,d;\mu,t)&=\lim_{\boldsymbol{q}\to0}
\frac{\mathrm{Im}\Pi_{jj}(\omega,\boldsymbol{q})}{\omega}
=
\begin{cases}
\mathrm{Re}\sigma_{\parallel}(\omega,d;\mu,t), &j=1,\\\\
\mathrm{Re}\sigma_{\perp}(\omega,d;\mu,t), &j\neq1,
\end{cases}
\label{Eq.9}
\end{align}
where $\omega$ is the photon frequency, $\mu$ measures the chemical potential with respect to the Dirac point, $j=1,2,...,d$ denotes the spatial component but the Einstein summation convention is not assumed. It is easy to verify that $\mathrm{Re}\sigma_{\chi}(\omega,d;\mu,t)$ respects the particle-hole symmetry, namely,
\begin{align}
\mathrm{Re}\sigma_{\chi}(\omega,d;\mu,t)&=\mathrm{Re}\sigma_{\chi}(\omega,d;-\mu,t)
=\mathrm{Re}\sigma_{\chi}(\omega,d;|\mu|,t),
\end{align}
and hence we are allowed to restrict our analysis to the case with $\mu\ge0$.

Hereafter, we focus on the interband (IB) LOCs
\begin{align}
\mathrm{Re}\sigma_{\chi}^{(\mathrm{IB})}(\omega,d;\mu,t)=\sum_{\kappa=\pm}\mathrm{Re}\sigma_{\chi}^{\kappa(\mathrm{IB})}(\omega,d;\mu,t).
\end{align}
The interband LOCs at the valley $\kappa$ can be explicitly written as
\begin{align}
\mathrm{Re}\sigma_{\chi}^{\kappa}(\omega,d;\mu,t) &
=\pi\int_{-\infty}^{+\infty}\frac{\text{d}^{d}\boldsymbol{k}}{(2\pi)^{d}}
\mathcal{F}_{\chi}^{\kappa;-+}(\boldsymbol{k})
\frac{f\left[\varepsilon_{\kappa}^{-}(\boldsymbol{k}),\mu\right]
-f\left[\varepsilon_{\kappa}^{+}(\boldsymbol{k}),\mu\right]}{\omega}
\delta\left[\omega-2v_{F}|\boldsymbol{k}|\right],
\end{align}
where $f(x,\mu)=\left\{1+\exp[\beta_{T}(x-\mu)]\right\}^{-1}$ represents the Fermi distribution function, and
\begin{align}
\begin{cases}
\mathcal{F}_{\parallel}^{\kappa;-+}(\boldsymbol{k})
=N_{f}e^{2}v_{F}^{2}\left(1-\frac{k_{1}k_{1}}{|\boldsymbol{k}|^2}\right),\\\\	 \mathcal{F}_{\perp}^{\kappa;-+}(\boldsymbol{k})
=N_{f}e^{2}v_{F}^{2}
\left[1-\frac{1}{d-1}\left(1-\frac{k_{1}k_{1}}{|\boldsymbol{k}|^2}\right)\right].
\end{cases}
\end{align}
with $N_{f}$ the flavor number of tilted Weyl fermions at one valley. In the following, we assume zero temperature $T\to0~\mathrm{K}$ for analytical calculation such that the Fermi distribution function reduces to the Heaviside step function $\Theta(\mu-x)$ which is defined as $\Theta(x)=0$ for $x\le0$ and $\Theta(x)=1$ for $x>0$.

The interband LOCs can be further expressed as
\begin{align}
\mathrm{Re}\sigma_{\chi}^{(\mathrm{IB})}(\omega,d;\mu,t)
&=S^{(\mathrm{IB})}(\omega,d)\Gamma_{\chi}^{(\mathrm{IB})}(\omega,d;\mu,t),
\end{align}
where the dimensional function
\begin{align}
S^{(\mathrm{IB})}(\omega,d)&=\left.\mathrm{Re}\sigma_{jj}^{(\mathrm{IB})}(\omega,d;\mu,t)\right|_{\mu=t=0}
=\frac{N_{f}\Omega_{d}}{(4\pi)^{d-1}}\frac{d-1}{d}\frac{\omega^{d-2}}{v_{F}^{d-2}}\sigma_{0},
\end{align}
with
$\sigma_0=e^2/4,$ and $\Omega_{d}=2\pi^{d/2}/\Gamma(d/2)$ the solid angle in $d$-D Euclidean space and $\Gamma(z)$ the Euler Gamma function. From this expression, we have a general scaling relation $S^{(\mathrm{IB})}(\omega,d)\propto (d-1)\omega^{d-2}$.
Specifically, one finds that the dimensional function $S^{(\mathrm{IB})}(\omega,d=1)=0$ results in $\mathrm{Re}\sigma_{jj}^{(\mathrm{IB})}(\omega,d=1;\mu,t)=0$, which explains the speciality of one spatial dimension.

\end{widetext}

Next, we will present the analytical expressions of the dimensionless function $\Gamma_{\chi}^{(\mathrm{IB})}(\omega,d;\mu,t)$ and joint density of state (JDOS) at zero temperature, and express $\Gamma_{\chi}^{(\mathrm{IB})}(\omega,d;\mu,t)$ and hence $\mathrm{Re}\sigma_{\chi}^{(\mathrm{IB})}(\omega,d;\mu,t)$ in terms of JDOS.

To better present the following results, we introduce two useful notations
\begin{align}		
\omega_{\pm}(t)&=\frac{2\mu}{|1\mp t|},\\
\zeta_{\pm}(\omega,t)&=\frac{\omega\pm2\mu}{\omega}\frac{\tilde{\Theta}(t)}{t},
\end{align}
and a more compact notation
\begin{align}
\tilde{\zeta}_{\pm}(\omega,t)=
	\begin{cases}
		+1, &\zeta_{\pm}(\omega,t)>+1,\\\\
		-1, &\zeta_{\pm}(\omega,t)<-1,\\\\
		\zeta_{\pm}(\omega,t), &-1\leq\zeta_{\pm}(\omega,t)\leq+1,
	\end{cases}
\end{align}
where $\tilde{\Theta}(x)=0$ for $x<0$ and $\tilde{\Theta}(x)=1$ for $x\geq0$.

After some straightforward algebra, the dimensionless function $\Gamma_{\chi}^{\mathrm{IB}}(\omega,d;\mu,t)$ can be expressed in a unified form as
\begin{align}
\Gamma_{\chi}^{(\mathrm{IB})}(\omega,d;\mu,t)
=\mathcal{G}_{\chi}^{(d)}\left[\tilde{\zeta}_{+}(\omega,t)\right]
+\mathcal{G}_{\chi}^{(d)}\left[\tilde{\zeta}_{-}(\omega,t)\right],
\end{align}
where the auxiliary function
\begin{align}
\begin{cases}
\mathcal{G}_{\parallel}^{(d)}(x)=Z_{d}~\frac{x}{2}~_{2}F_{1}\left(\frac{1}{2},\frac{1-d}{2};\frac{3}{2};x^{2}\right),& d\ge 0,\\\\
\mathcal{G}_{\perp}^{(d)}(x)=\frac{d}{d-1}~\mathcal{G}_{\parallel}^{(d-2)}(x)-\frac{1}{d-1}\mathcal{G}_{\parallel}^{(d)}(x),&d\ge 2,
\end{cases}
\end{align}
where
\begin{align}
Z_{d}=\frac{1}{~_{2}F_{1}(\frac{1}{2},\frac{1-d}{2};\frac{3}{2};1)}
=\frac{2~\Gamma\left(\frac{d+2}{2}\right)}{\sqrt{\pi}~\Gamma\left(\frac{d+1}{2}\right)},
\end{align}
with $_{2}F_{1}(\alpha,\beta;\gamma;x)$ the Gauss hypergeometric function.

It is emphasized that $\mathcal{G}_{\parallel}^{(d)}(x)$ is well defined only when $d\ge0$, while $\mathcal{G}_{\parallel}^{(d)}(x)$ when $d\ge2$. Physically, these corresponds to that the parallel component $\chi=\parallel$ can always be defined if $d\ge0$, and that the perpendicular component $\chi=\perp$ can only be defined when $d\ge2$ since the parallel component must be previously defined as a reference. Evidently, $\mathcal{G}_{\chi}^{(d)}(x)$
is an odd function always satisfying $\mathcal{G}_{\chi}^{(d)}(0)=0$ and $\mathcal{G}_{\chi}^{(d)}(\pm1)=\pm\frac{1}{2}$ for $d\ge0$. All of these expressions are valid for both doped cases ($\mu=0$ and $\mu\neq 0$), all tilted phases ($t=0$, $0<t<1$,
$t=1$, and $t>1$), both spatial components ($\chi=\parallel$ and $\chi=\perp$). In addition, they restore analytically
to all the previous results in 1D \cite{arXivHou2022}, 2D \cite{PRLCarbotte2006,PRBGusynin2007,PRLMikhailov2007,PRLMarel2008,
PRLMak2008,PRBStauber2008,JPSJNishine2010,PRBVerma2017,PRBHerrera2019,PRBGoerbig2019,PRBTan2021,PRBTan2022,PRBWild2022,
arXivHou2022}, and 3D \cite{PRBAshby2014,PRBCarbotte2016,PRBCarbotte2017,arXivHou2022}.

Interestingly, for $\omega=2\mu$ and $t>0$, we have
\begin{align}
\tilde{\zeta}_{-}(\omega=2\mu,t>0)&=0,\\
\tilde{\zeta}_{+}(\omega=2\mu,t>0)&=
		\tilde{\Theta}(2-t)\Theta(t)+\frac{2\Theta(t-2)}{t}.
\end{align}
As a consequence, these is always a robust fixed point
\begin{align}
\hspace{-0.3cm}\Gamma_{\chi}^{(\mathrm{IB})}(\omega=2\mu,d\ge 2;\mu,0<t\le 2)
=\mathcal{G}_{\chi}^{(d)}\left(+1\right)=\frac{1}{2},
\end{align}
for $\chi=\parallel,\perp$ when $0<t\le 2$, which means that the robust fixed point at $\omega=2\mu$
is universal for tilted Dirac bands in arbitrary spatial dimensionality with $d\ge 2$ rather than only in 2D and 3D.
It is noted that this behavior was also reported previously in 2D and 3D tilted Dirac bands \cite{PRBTan2022,PRBCarbotte2016}, the underlying physical reason of which was intuitively presented in terms of the geometric structures of Fermi surface and energy resonance contour in Ref.\cite{arXivHou2022}.

Utilizing the definition of $\Gamma_{\perp}^{(\mathrm{IB})}(\omega,d;\mu,t)$, we obtain a recurrence relation
\begin{align}
&\Gamma_{\perp}^{(\mathrm{IB})}(\omega,d;\mu,t)
\nonumber\\&
=\frac{d}{d-1}\Gamma_{\parallel}^{(\mathrm{IB})}(\omega,d-2;\mu,t)
-\frac{\Gamma_{\parallel}^{(\mathrm{IB})}(\omega,d;\mu,t)}{d-1}.
\label{Gammaperp}
\end{align}
when $d\ge2$.

It is helpful to express $\Gamma_{\chi}^{(\mathrm{IB})}(\omega,d;\mu,t)$ and $\mathrm{Re}\sigma_{\chi}^{(\mathrm{IB})}(\omega,d;\mu,t)$ in terms of JDOS.
By definition, the JDOS at zero temperature
\begin{align}
J(\omega,d)
&=\frac{N_{f}}{2}\sum_{\kappa=\pm}\int\frac{\mathrm{d}^{d}\boldsymbol{k}}{(2\pi)^{d}}\left\{ \Theta\left[\mu-\varepsilon_{-}^{\kappa}(\boldsymbol{k})\right]
\right.\nonumber\\&\left.\hspace{1.5cm}
-\Theta\left(\mu-\varepsilon_{+}^{\kappa}(\boldsymbol{k})\right)\right\}\delta\left[\omega-2v_{F}|\boldsymbol{k}|\right].
\end{align}

Interestingly, when $d\geq2$, the JDOS satisfies the relation
\begin{align}
\frac{J(\omega,d)}{J_{0}(\omega,d)}&=
\mathcal{G}_{\parallel}^{(d-2)}\left[\tilde{\zeta}_{+}(\omega)\right]
+\mathcal{G}_{\parallel}^{(d-2)}\left[\tilde{\zeta}_{-}(\omega)\right]
\nonumber\\&
=\Gamma_{\parallel}^{(\mathrm{IB})}(\omega,d-2;\mu,t),
\label{Gammadm2}
\end{align}
with
\begin{align}
J_{0}(\omega,d)&=\frac{N_{f}\Omega_{d}}{(4\pi)^{d}}\frac{\omega^{d-1}}{v_{F}^{d}},
\end{align}
which is equivalent to
\begin{align}
\Gamma_{\parallel}^{\mathrm{(IB)}}(\omega,d;\mu,t)=\frac{J(\omega,d+2)}{J_{0}(\omega,d+2)},
\label{Gammad}
\end{align}
for $d\ge 0$.

Substituting Eqs.(\ref{Gammadm2}) and (\ref{Gammad}) into Eq.(\ref{Gammaperp}), one can easily express $\Gamma_{\perp}^{\text{(IB)}}(\omega,d;\mu,t)$ as
\begin{align}
&\Gamma_{\perp}^{\text{(IB)}}(\omega,d;\mu,t)
\nonumber\\&
=\frac{d}{d-1}\frac{J(\omega,d)}{J_{0}(\omega,d)}-\frac{1}{d-1}\frac{J(\omega,d+2)}{J_{0}(\omega,d+2)},
\label{Gammadperp}
\end{align}
for $d\ge2$. Hence, we show the fact that $\Gamma_{\chi}^{\text{(IB)}}(\omega,d;\mu,t)$ can be expressed by JDOS.
Moreover, by multiplying $S^{(\mathrm{IB})}(\omega,d)$ to both sides of Eqs.(\ref{Gammad}) and (\ref{Gammadperp}),
one has the interband LOCs. The parallel component of the interband LOC $\mathrm{Re}\sigma_{\parallel}^{(\mathrm{IB})}(\omega,d;\mu,t)$ for $d\ge 0$ takes
\begin{align}
\mathrm{Re}\sigma_{\parallel}^{(\mathrm{IB})}(\omega,d;\mu,t)=\frac{32\pi^{2}v_{F}^{4}\sigma_{0}}{\omega^{3}}(d-1)J(\omega,d+2),
\end{align}
which accounts for the vanishing of $\mathrm{Re}\sigma_{\parallel}^{(\mathrm{IB})}(\omega,d;\mu,t)$ in one spatial dimension. The perpendicular component of the interband LOC $\Gamma_{\perp}^{(\mathrm{IB})}(\omega,d;\mu,t)$ for $d\ge 2$ as
\begin{align}
&\mathrm{Re}\sigma_{\perp}^{(\mathrm{IB})}(\omega,d;\mu,t)
\nonumber\\&
=\frac{4\pi v_{F}^{2}\sigma_{0}}{\omega}\left[J(\omega,d)-\frac{8\pi v_{F}^{2}}{\omega^{2}}J(\omega,d+2)\right].
\end{align}

In conclusion, we found that the unified analytical form of the LOCs in tilted Weyl fermions for arbitrary spatial dimensionality, which can be expressed in terms of the JDOS. They are valid for both undoped and doped cases, both
parallel and perpendicular components, and all the tilted phases. Additionally, they restore to the analytical results
of previous works in 1D, 2D, and 3D. The robust fixed point $\Gamma_{\chi}^{(\mathrm{IB})}(\omega=2\mu,d\ge 2;\mu,0<t\le 2)=1/2$ is universal for tilted Dirac bands in arbitrary spatial dimensionality. Our work offers a once-for-all method prior to the one-by-one calculation of the LOCs in tilted Weyl fermions and deep insights into the impacts of dimensionality therein.

This work is partially supported by the National Natural Science Foundation of China under Grant Nos.11547200 and the Sichuan Science and Technology under Grant No.2022-YCG057.

\end{document}